\documentclass[useAMS,usenatbib]{mn2e}

\usepackage{amsmath}
\usepackage{graphicx}
\usepackage{amssymb}

\newcommand{\x}{\mathbf{x}}
\newcommand{\xe}{\mathbf{x_\mathrm{ext}}}
\newcommand{\xr}{\mathbf{x_\mathrm{ran}}}

\newcommand{\nside}{\textsc{\small{NSIDE}}}

\title[CMB polarization for probing the Cold Spot]
{CMB polarization as a probe of the anomalous nature of the Cold Spot}
\author[P. Vielva et al.]
{P. Vielva$^{1}$, E. Mart{\'\i}nez-Gonz\'alez$^{1}$, M. Cruz$^{2}$, R.B. Barreiro$^{1}$, M. Tucci$^{1}$ \\
$^{1}$Instituto de F{\'\i}sica de Cantabria (CSIC - Univ. de Cantabria), Avda. Los Castros s/n, 39005 - Santander, Spain\\
$^{2}$Departamento de Matem\'aticas, Estad{\'\i}stica y Computaci\'on, Univ. de Cantabria, Avda. Los Castros s/n, 39005 - Santander, Spain\\
\hspace{0.1cm}E-mails : vielva@ifca.unican.es, martinez@ifca.unican.es, cruz@ifca.unican.es, barreiro@ifca.unican.es, mtucci72@gmail.com
\\}
\date{Accepted ???. Received ???; in original form \today}
\pagerange{\pageref{firstpage}--\pageref{lastpage}}
\pubyear{1998}

\begin{document}
\maketitle
\label{firstpage}

\begin{abstract}
One of the most interesting explanations for the non-Gaussian Cold Spot detected in the WMAP data by~\cite{vielva04}, is that it arises from the interaction of the CMB radiation with a cosmic texture~\citep{cruz07b}. In this case, a lack of polarization is expected in the region of the spot, as compared to the typical values associated to large fluctuations of a Gaussian and isotropic random field.
In addition, other physical processes related to a non-linear evolution of the gravitational field could lead to a similar scenario.
However, some of these alternative scenarios (e.g., a large void in the large scale structure) have been
shown to be very unlikely.
In this work we characterise the polarization properties of the Cold Spot under both hypotheses: a large Gaussian fluctuation and an anomalous feature generated, for instance, by a cosmic texture. We also propose a methodology to distinguish between them, and we discuss its discrimination power as a function of the instrumental noise level. In particular, we address the cases of current experiments, like WMAP and Planck, and others in development as QUIJOTE. We find that for an ideal experiment with a high polarization sensitivity, the Gaussian hypothesis could be rejected at a significance level better than 0.8\%. While WMAP is far from providing useful information in this respect, we find that Planck will be able to reach a significance of around 7\%; in addition, we show that the ground-based experiment QUIJOTE could provide a significance of around 1\%, close to the ideal case.
If these results are combined with the significance level found for the Cold Spot in temperature, the capability of QUIJOTE and Planck to reject the alternative hypothesis becomes 0.025\% and 0.124\%, respectively.
\end{abstract}


\section{Introduction}
\label{intro}

Cosmology is living a golden age thanks to the analysis of high quality data that are being collected during the last years by several experiments. Among the observables used to probe the nature of the Universe, the Cosmic Microwave Background (CMB) temperature and polarization fluctuations provide a unique tool that is helping to establish a well defined picture of the origin, evolution, and matter and energy content of the universe
~\citep[e.g.,][see \citealt{barreiro10} for a recent review]{komatsu09,gupta09}.
However, since the public release of the WMAP 1st-year data in 2003~\citep{bennett03}, and the subsequent data releases~\citep[][]{spergel07,hinshaw09,jarosik10}, several results have been reported that seem to challenge the statistically isotropic and Gaussian nature of the CMB, predicted by the standard inflationary theory.

Among these anomalies, the exceptionally large and cold spot (hereinafter the Cold Spot or CS) that was identified in the southern 
hemisphere (l = $209^\circ$, b = $57^\circ$) through a wavelet analysis~\citep{vielva04,cruz05} is one of the features that has attracted
more attention from the scientific community. The CS has been widely confirmed by 
subsequent analyses~\citep[e.g.,][]{mukherjee04,cayon05,mcewen05,rath07,vielva07,gurzadyan09,pietrobon08,rossmanith09}, 
carried out by different groups and using different kinds of techniques.

Recently~\cite{zhang10} have claimed that the Cold Spot originally found by~\cite{vielva04} was, actually, an artifact
caused by the particular choice of the Spherical Mexican Hat Wavelet (SMHW) as the tool to analyse the data. To support this argument, the authors showed how the use of isotropic filters with variable width, like a top-hat or a Gaussian function, failed to provide a deviation from Gaussianity. We do not agree with the conclusions reached in that paper. The results obtained by the authors just indicate that not all filtering
kernels are equally optimal to detect or amplify a particular signature. In particular, wavelets (which are compensated filters) are better suited for this purpose than other non-optimised kernels. It is well known that wavelets increase the signal-to-noise ratio 
of those features with a characteristic scale similar to the one of the wavelet. This amplification is obtained by filtering out the 
instrumental noise and the inflationary CMB fluctuations at smaller and larger scales. The arguments given by~\cite{zhang10} have been merely repeated by~\cite{bennett10} in a recent work.

A number of possible explanations for the CS have been suggested in the literature,
namely contamination from residual foregrounds~\citep[e.g.,][]{liu05,coles05}, particular brane-world models~\citep{cembranos08}, the collision of cosmological bubbles~\cite[e.g.][]{chang09}
the non-linear integrated
Sachs-Wolfe effect produced by the large scale structure~\citep[e.g.,][]{tomita05,inoue06,rudnick07,garciabellido08,masina09}, or inverse 
Compton scattering via the Sunyaev-Zeldovich effect, supported by the presence of a large cluster of galaxies in the direction of
the CS~\citep[the Eridanus super-group,][]{brough06}.
However, some works have shown that these explanations are very unlikely~\citep[e.g.,][]{cruz06,cruz08,smith10}, since, depending on the case, they would require very special conditions to be able to explain the CS, such as 
a very particular mixing-up of the foreground emissions, an unfeasible electron gas distribution, a very peculiar situation
of the Milky Way with respect to some hypothetical large voids, or the existence of huge voids much larger than the ones expected from 
the standard structure formation scenario. In particular, the latter would imply a much more extreme departure from Gaussianity
than the one that these models are trying to explain!

Nevertheless, there is an alternative hypothesis that has not been ruled out yet, which is compatible with
current observations.~\cite{cruz07b} suggested that the CS 
could be produced by the non-linear evolution of the gravitational potential generated by a collapsing cosmic texture. In that work, a Bayesian
analysis showed that the texture hypothesis was preferred with respect to the pure standard Gaussian scenario, and that the values describing the
properties of the texture were compatible with current cosmological observations. In particular, 
the energy scale for the symmetry breaking that generates
this particular type of topological defect ($\phi_0 = 8.7\times 10^{15}$ GeV), was in agreement with the upper limits established by means of the 
angular power spectrum~\citep[e.g.,][]{bevis04,urrestilla08}.

Of course, this result does not guarantee by itself the existence of cosmic textures, nor that the CS is caused
by a collapsing texture. In fact, further tests are needed, and some of them were already indicated in~\cite{cruz07b}. First, the texture model makes predictions about the expected
number of cosmic textures with an angular scale equal or greater than $\theta$. In particular,  
the presence of around 20 cosmic 
textures with $\theta \gtrsim 1^\circ$ is predicted. Some works, like~\cite{gurzadyan09,vielva07,pietrobon08},
have already reported the existence of other anomalous spots, which could potentially be related to the presence of additional textures. Second, the pattern of the
CMB lensing signal induced by such a texture is known, and high resolution CMB experiments (like the Atacama Cosmology Telescope and the South Pole Telescope)
should be able to detect such a signal, if present. This issue has recently been addressed by~\cite{das09}. Finally, the polarization of the CMB is an
additional source of information that provides further insight on the texture hypothesis. 
A lack of polarization is expected for the texture hypothesis, as 
compared to the typical values associated to large fluctuations of a 
Gaussian and isotropic random field. This is because the effect of a 
collapsing texture on the CMB photons is merely gravitational. This 
difference in the polarization is the topic of this work.

Nevertheless, it is worth recalling that a collapsing texture is not the only way of producing
a local non-linear evolution of the gravitational potential, and, therefore, a relative lack of the local polarization
signal. Other physical processes could also generate such a secondary
anisotropy on the CMB photons. In fact, some of these effects have also been proposed as
possible explanations for the CS. For instance, as previously mentioned, a very large void~\citep[e.g.,][]{rudnick07}
could produce the required non-linear evolution and, therefore, it would be affected by a relative lack of polarization.
However, this explanation is discarded from both current large scale structure modelling~\citep[e.g.,][]{cruz08,smith10} and dedicated observations~\citep{granett09,bremer10}. For this reason, in this paper we consider the non-linear Integrated Sachs-Wolfe (also called Rees-Sciama) effect caused by a collapsing texture as the most
plausible explanation.
In any case, we remark that the results derived in this paper for the texture model can also be expected in the
most general situation of
any physical process producing
CMB secondary anisotropies, in the form of large spots in temperature, via the non-linear Sachs-Wolfe effect.

The paper is organised as follows. In Section~\ref{sec:chara} we provide a characterisation of the radial profile (both in temperature and polarization)
for Gaussian spots as extreme as the CS. For comparison, we also investigate the case of random positions. A method, which exploits the correlation between the temperature and the polarization profiles, is proposed to discriminate between the Gaussian (null) and the texture (alternative) hypotheses in Section~\ref{sec:method}. The results are given in Section~\ref{sec:results},
where the ability to discriminate between the two considered hypotheses is discussed for different instrumental sensitivities. Finally, conclusions are presented in Section~\ref{sec:final}.

\begin{figure*}
\includegraphics[width=8cm,keepaspectratio]{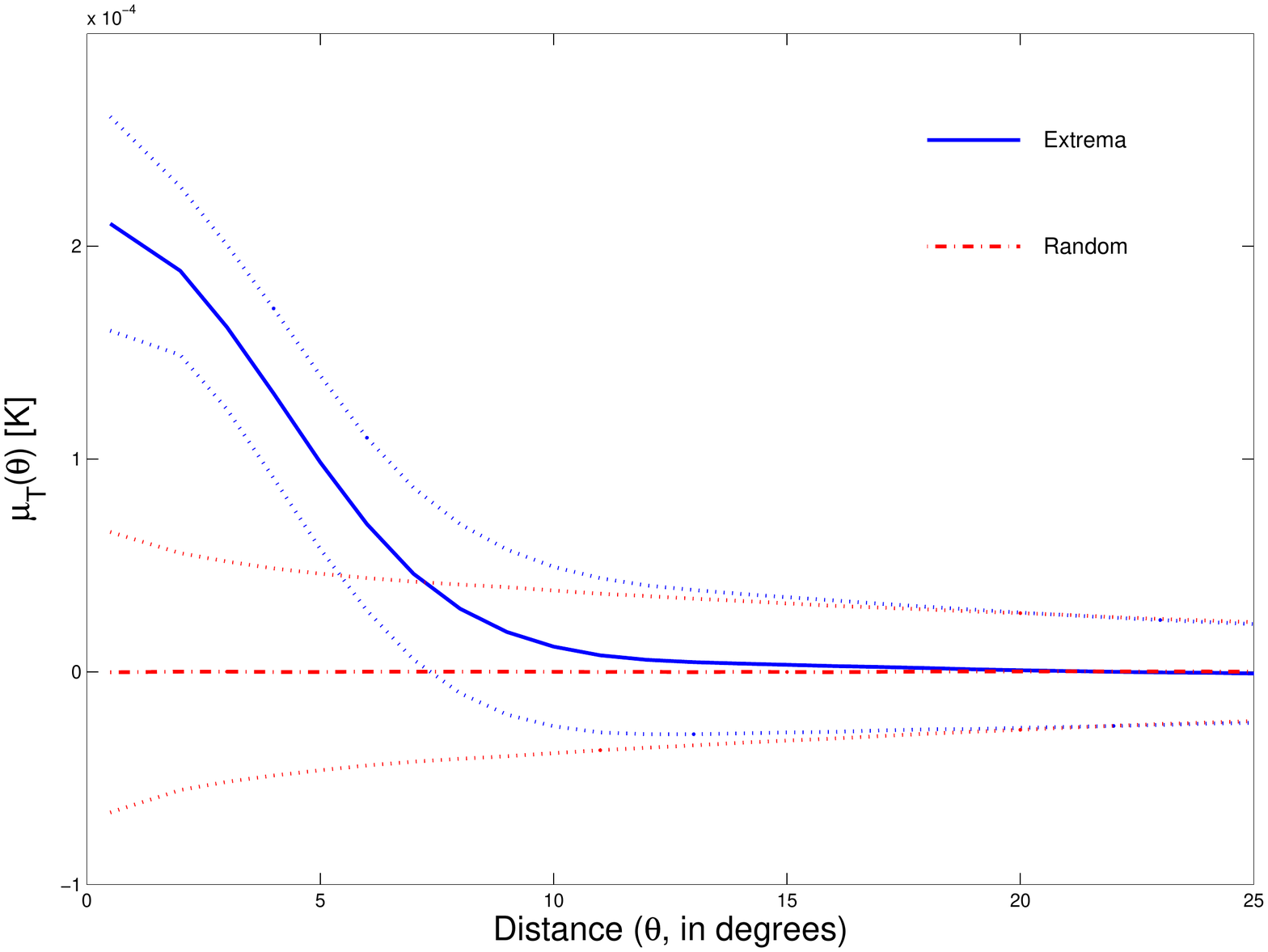}
\includegraphics[width=8cm,keepaspectratio]{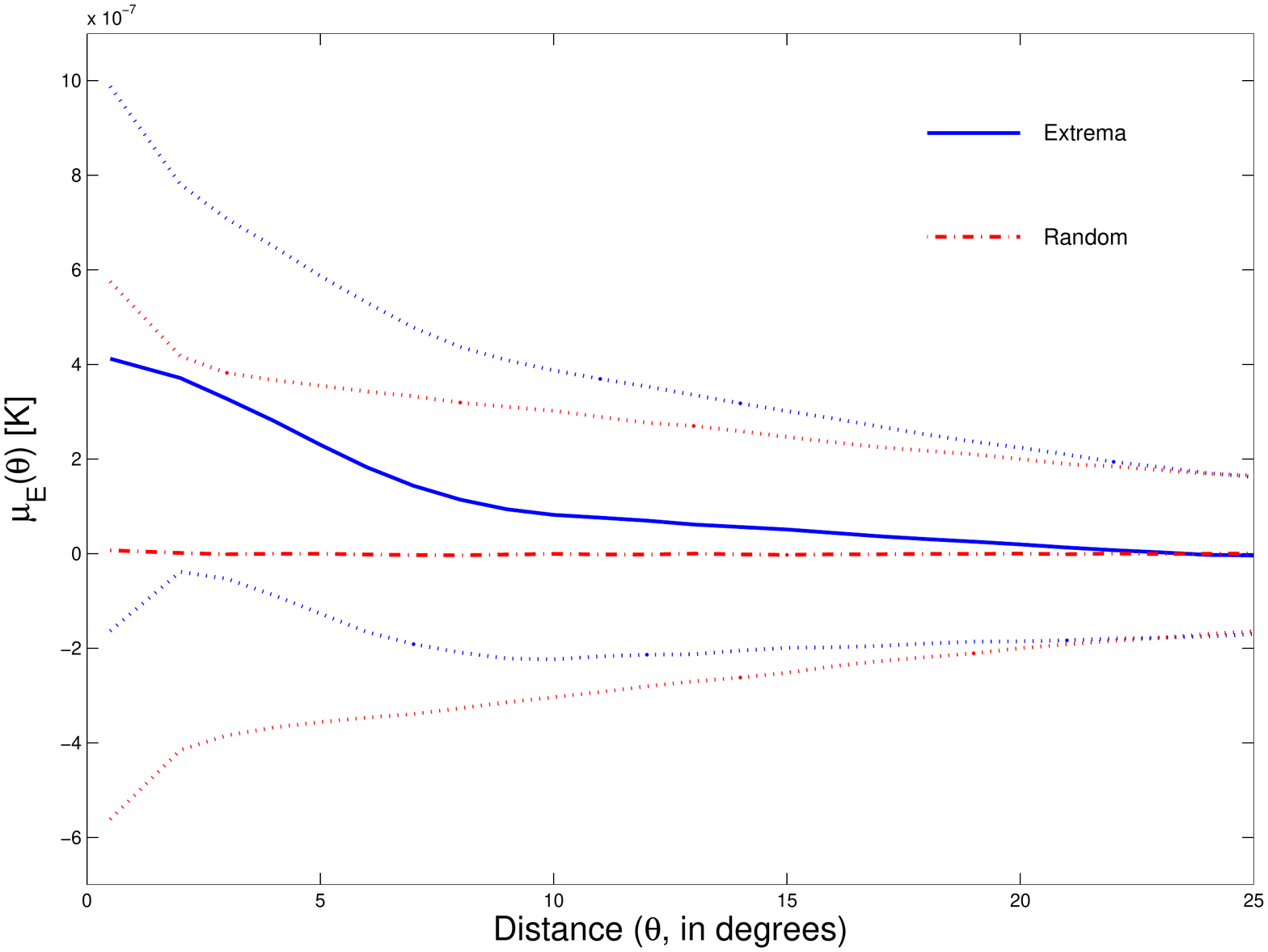}
\caption{\label{fig:profiles} Mean temperature ($\mu_T(\theta)$, left panel) and E-mode polarization ($\mu_E(\theta)$, right panel) radial profiles at \emph{Extrema} positions, with an amplitude in the temperature maps, at least, as large as the one of the CS (solid blue lines), and at \emph{Random} positions in the temperature maps (dot-dashed red lines). The dotted lines show their corresponding dispersion. See text for details.}
\end{figure*}

\section{Characterisation of the temperature and polarization signals}
\label{sec:chara}

As already mentioned, the cross-correlation of the temperature ($T$) and polarization\footnote{We do not consider the B-mode of polarization since current observations~\citep[e.g.,][]{gupta09} show that this is significantly lower than the E-mode.} ($E$) signals around the position of the CS, could be an excellent discriminator between the null and alternative hypotheses. In other words, this quantity could indicate whether this feature is better described by a standard Gaussian and isotropic field, or, conversely, by a non-standard cosmological model producing temperature spots which do not present a correlated polarization feature (as the topological defects). In this latter case, the CS is assumed to be caused by a secondary anisotropy of the CMB photons, altered by a
non-linearly evolving gravitational potential produced, for instance, by a collapsing cosmic texture.

Hence, this \emph{alternative} hypothesis ($H_1$) would correspond to CMB fluctuations generated by the standard inflationary model, but with a non-negligible contribution from topological defects (as it would be the case for the CS). Conversely, the \emph{null} hypothesis ($H_0$) would be the case in which all the CMB fluctuations (including the CS) are due to a pure standard Gaussian and isotropic field. It is interesting to point out that for the case of the alternative hypothesis, the E-mode signal is not expected to contain contributions from scalar perturbations but only from vector perturbations, which are around one order of magnitude smaller. Therefore, for a CMB temperature feature as extreme as the CS, one would expect more polarization signal if such 
temperature fluctuation is caused by the standard inflationary model, than for the case in which, for instance, 
a collapsing cosmic texture is producing such a large spot.

We aim to characterise the CMB temperature and (E-mode) polarization features through a radial profile. The reason to adopt this characterisation is simple: the shape of the CS is close to spherical, with a typical size of around 10 degrees~\citep{cruz06}. 

At this point, it is important to recall that the CS was first identified as an anomalous feature with
an amplitude of $-4.57$ times the dispersion of the Spherical Mexican Hat Wavelet (SMHW) coefficients at a wavelet scale $R = 250$ 
arcmin~\citep[for details see][]{vielva04}. Follow-up tests explored additional characteristics of the CS finding even lower p-values~\citep[e.g.,][]{cruz05}, but for the sake of simplicity and robustness, we adopt the original detection as the statistical property that characterises the CS.

Let us define, for a given position $\x$, the radial profile in temperature $\mu_T(\x,\theta)$ and in polarization $\mu_E(\x,\theta)$ as:
\begin{eqnarray}
\mu_T\left( \x, \theta \right) & = & \frac{1}{N} 
\sum_{ \x^*}T\left(\x^*\right) \\
\mu_E\left( \x, \theta \right) & = & \frac{1}{N} 
\sum_{ \x^*}E\left(\x^*\right), 
\label{eq:correl_def}
\end{eqnarray}
where $T$ and $E$ are the temperature and the E-mode polarization maps, respectively. The sums are extended over the positions $\x^*$ 
which are at a distance $\theta^*$ from $\x$ --i.e., $\theta^* \equiv \arccos\left( \x^* \cdot \x \right)$-- such as:
$\theta^*  \in \left[\theta-\Delta\theta/2, \theta+\Delta\theta/2 \right]$. $\Delta\theta$ is the width of the considered rings and $N$ represents the number of positions (or pixels in a map at a given angular resolution) satisfying the previous condition.

In Figure~\ref{fig:profiles} we plot the mean value and dispersion of the temperature and polarization radial profiles for two
different cases. The first case, labelled as \emph{Extrema}, corresponds to the radial profiles $\mu_T(\x,\theta)$ and $\mu_E(\x,\theta)$ associated to positions $\xe$ where the CMB Gaussian temperature field has a feature, at least, as extreme as the CS (i.e., having an amplitude above 4.57 times the dispersion of the wavelet coefficients at a wavelet scale $R = 250$ arcmin, in absolute value). Note that, although the CS is actually cold (i.e, it is a minimum), in this work we will consider the more general case of having an extremum of the CMB field. We adopt this criterion since, for the case of cosmic textures, either hot or cold spots can be produced. The second case is labelled as \emph{Random}, and it corresponds to the radial profiles associated to random positions $\xr$ selected in the CMB Gaussian temperature field. 

These mean radial profiles have been obtained after averaging over many simulations, carrying out the following procedure. First, a CMB Gaussian simulation is generated (containing T, Q, and U maps) at a resolution given by the HEALPix~\citep{gorski05}
parameter $\nside=64$. Subsequently, the temperature component of the simulation is filtered with the SMHW at a scale $R=250$ arcmin. A feature as extreme as the CS is then sought in the wavelet coefficient map. If this is not found, a new simulation is generated. Conversely, if a CS-like feature is present in the temperature map, we compute the E map from the pseudo-scalars Q and U, as well as the temperature $\mu_T(\xe,\theta)$ and polarization $\mu_E(\xe,\theta)$ profiles at the position $\xe$ where the extremum is located. In addition, a random position in the temperature map is selected and the $\mu_T(\xr,\theta)$ and
$\mu_E(\xr,\theta)$ profiles are computed at this random position $\xr$. 

The left panel in Figure~\ref{fig:profiles} shows both cases (\emph{Extrema} as the blue solid line, and \emph{Random} as the green dot-dashed line) for temperature,
while the right panel corresponds to the (E-mode) polarization. Let us remark that, for \emph{Extrema} that are cold spots, the absolute value of the
profile has been considered.
The curves show the profiles from 0.5 to 25 degrees, with a step of $\Delta\theta = 0.5$ degrees.
We also plot the 1-$\sigma$ level (dotted lines) associated to the probability distribution of the profiles at a given distance, obtained from the 10000 simulations used to compute these estimates.
Note that, for the case of the polarization signal, the \emph{Extrema} and the \emph{Random} profiles overlap at the 1-$\sigma$ level, which
indicates that very little information can be obtained from the analysis of the polarization signal alone. This is a justification to consider the
polarization information only via the cross-correlation with the temperature fluctuations.
Hence, the differences between these curves, expressed in terms of their mutual correlations, are the ingredients used to define a methodology to discriminate between the
standard Gaussian (null) and the non-standard \emph{cosmic texture} (alternative) hypotheses.
This is addressed in the next Section.

\section{The methodology}
\label{sec:method}

In this section we describe a methodology to distinguish between the competitive hypotheses already mentioned: the standard Gaussian and isotropic inflationary model ($H_0$, null hypothesis),
and a non-standard model that accounts, for instance, for cosmic textures, in addition to CMB fluctuations coming from the standard inflationary model ($H_1$, alternative hypothesis).

The key point to discriminate between these two scenarios is to exploit the differences between the cross-correlation of the temperature $\mu_T(\theta)$ and the polarization $\mu_E(\theta)$ radial profiles described in the previous Section.

Let us define, first, these two hypotheses (Sections~\ref{subsec:signal_h0} and~\ref{subsec:signal_h1}) in terms of the temperature and polarization profiles. Afterwards (Section~\ref{subsec:discriminator}) we will build the discriminator, based on the Fisher discriminant.

\subsection{The correlation signal for the $H_0$ hypothesis}
\label{subsec:signal_h0}

Under the assumption of the null hypothesis $H_0$ (i.e., a CMB signal completely described in terms of the standard inflationary model), the cross-correlation of the temperature and polarization profiles at position $\xe$ (i.e., where the CMB  
temperature map presents a CS-like feature) is given by:
\begin{equation}
\label{eq:cross_h0}
\xi_{H_0} \left( i \right) \equiv  \mu_T\left(\xe, \theta_T\right) \mu_E\left(\xe, \theta_E\right).
\end{equation}
$\xi_{H_0}$ is a vector with $n_c = \theta_T \times \theta_E$ components (i.e., $i = \left\lbrace 1, 2, ..., n_c \right\rbrace$). In our analysis we have $n_c = 324$ since we consider values 
of $\theta_T$ and $\theta_E$ from 1 to 18 degrees, with $\Delta \theta = 1$ degree. We have tested that including smaller or larger scales does not increase significantly the discrimination ability of our estimator.
After averaging over simulations, we can obtain both, the mean value of this signal vector 
($\bar{\xi}_{H_0}$), and the covariance matrix among the different components of the vector ($C_{H_0}$, with dimension $n_c \times n_c$). We define the $i$-component of the signal vector and the $(j,k)$-element of the covariance matrix as:
\begin{eqnarray}
\label{eq:mean_h0}
\bar{\xi}_{H_0}\left(i\right) & = & \frac{1}{N_s}\sum_{n=1}^{N_s} \xi_{H_0,n}\left(i\right), \\
\label{eq:cov_h0} C_{H_0}\left(j, k\right) & = & \frac{1}{N_s}\sum_{n=1}^{N_s} \big( \xi_{H_0,n}\left(j\right) - \bar{\xi}_{H_0}\left(j\right)\big) \times \\
& & ~~~~~~~~~~\big( \xi_{H_0,n}\left(k\right) - \bar{\xi}_{H_0}\left(k\right)\big) \nonumber,
\end{eqnarray}
where $N_s$ is the total number of simulations used to compute these estimators. In our analysis, we consider $N_s$ = 10,000.

\begin{figure*}
\includegraphics[width=5.5cm,keepaspectratio]{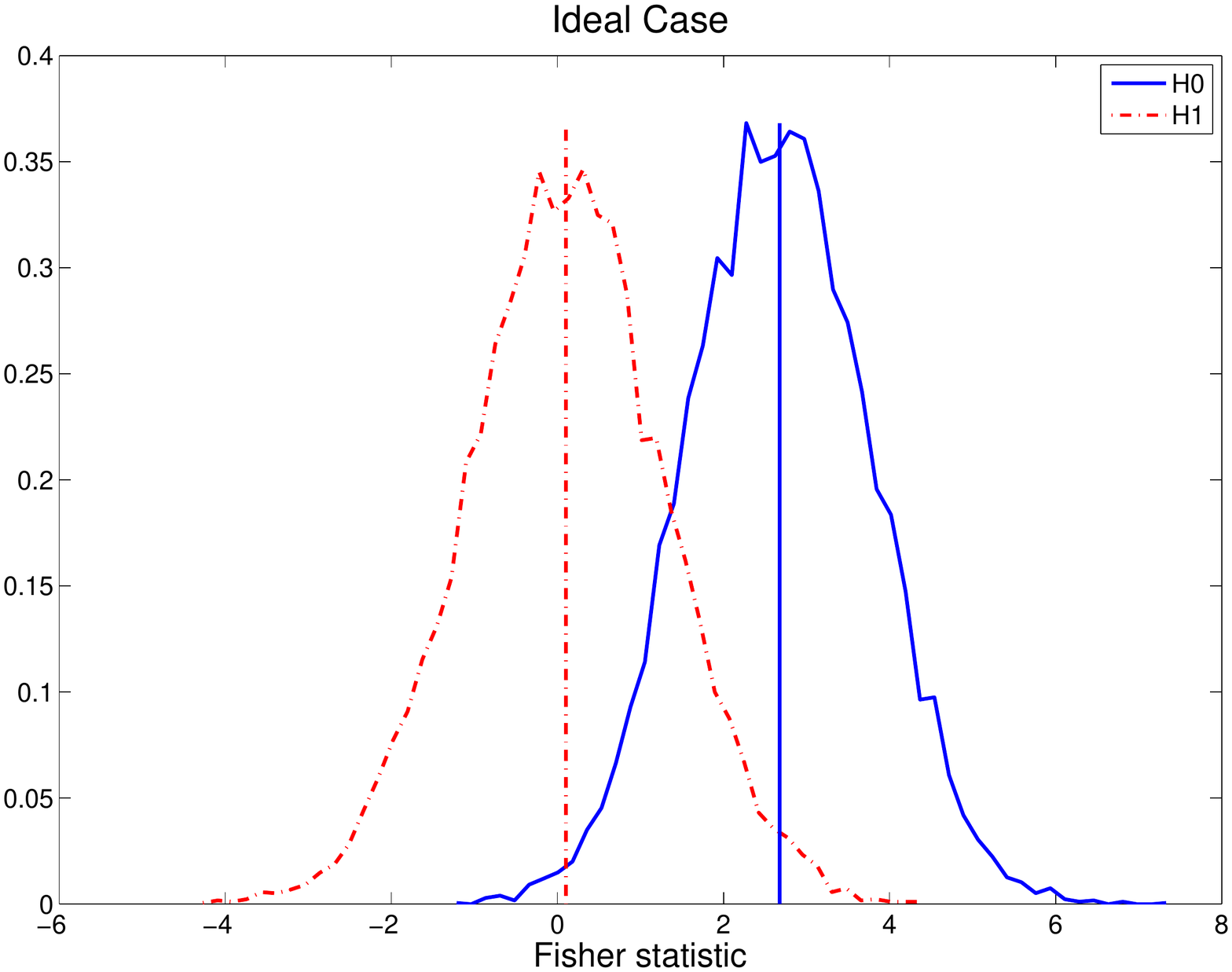}
\includegraphics[width=5.5cm,keepaspectratio]{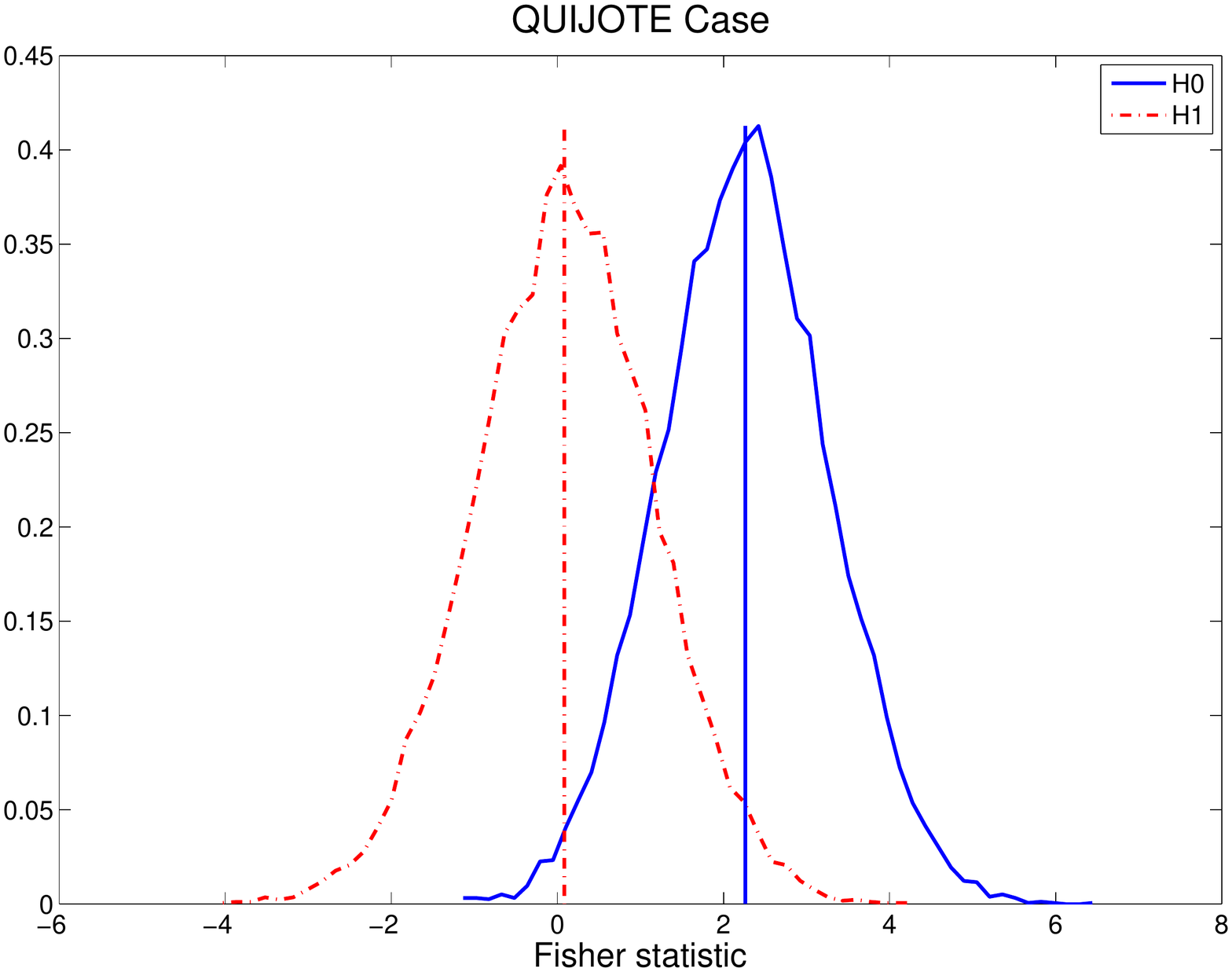}
\includegraphics[width=5.5cm,keepaspectratio]{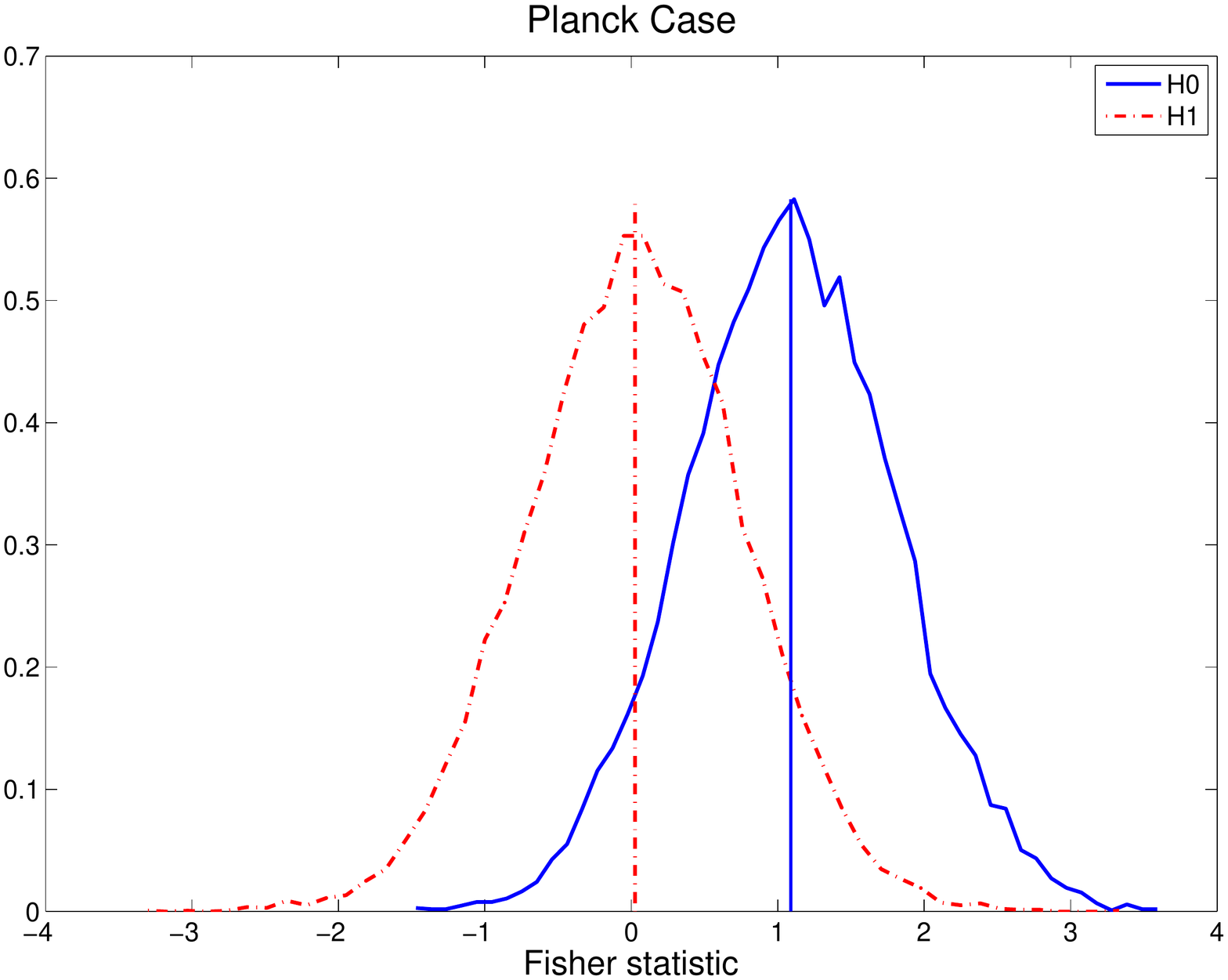}
\caption{\label{fig:fisher}Fisher discriminants for an ideal noise-free experiment, the QUIJOTE telescope and the ESA Planck satellite (from left to right). Blue solid lines correspond to the Fisher discriminant $\tau_{H_0}$ for the null ($H_0$) hypothesis, whereas the Fisher discriminant $\tau_{H_1}$ for the alternative ($H_1$) hypothesis is shown as red dot-dashed lines. Significance levels (at a power of the test of 0.5) are: 0.008, 0.014 and 0.069, respectively.}
\end{figure*}

\subsection{The correlation signal for the $H_1$ hypothesis}
\label{subsec:signal_h1}

Under the assumption of the alternative hypothesis $H_1$ (e.g., the case in which the CMB fluctuations are generated from the standard Gaussian and isotropic field, plus a contribution of cosmic textures which is, indeed, responsible for the CS), the cross-correlation of the temperature and polarization profiles at position $\xe$ (where the CMB temperature map has a feature as extreme as the CS) is given by:
\begin{equation}
\label{eq:cross_h1}
\xi_{H_1} (i) \equiv  \mu_T\left(\xe, \theta_T\right) \mu_E\left(\xr, \theta_E\right) + \beta_{TE},
\end{equation}
where the first term at the right-hand side of the equation corresponds to the correlation between a radial profile in temperature for a CS-like feature, and a radial profile in polarization associated to a typical fluctuation generated by the Gaussian and isotropic component. This term accounts for the fact that a cosmic texture would add an almost negligible polarization 
signal. As already mentioned, the reason is that textures do not produce
E-mode scalar perturbations, but vector perturbations, which are one order of magnitude smaller than the former.
In addition, the term $\beta_{TE}$ is a small correction (as compared to the previous term) that can be seen as a bias accounting from the residual correlations between the temperature and polarization profiles in a random position $\xr$ of
the CMB $T$ map, i.e., $\beta_{TE} \equiv \bar{\xi}_{H_0}^{\xr}$, where:
\begin{equation}
\label{eq:bias1}
\bar{\xi}_{H_0}^{\xr}\left(i\right) = \frac{1}{N_s}\sum_{n=1}^{N_s} \xi_{H_0, n}^{\xr}\left(i\right),
\end{equation}
with 
\begin{equation}
\label{eq:bias2}
\xi_{H_0}^{\xr} (i) \equiv  \mu_T\left(\xr, \theta_T\right) \mu_E\left(\xr, \theta_E\right).
\end{equation}
Note that the bias term $\beta_{TE}$ is required to account for the typical correlations that exist between the temperature and the polarization field.
In other words, it accounts for the TE cross-correlations due to the underlying isotropic and Gaussian fluctuations, where the CS (caused by the
cosmic texture) is placed. 

As in the previous case, $\xi_{TE}\left(H_1\right)$ is a vector with $n_c = \theta_T \times \theta_E$ components. Its mean value ($\bar{\xi}_{H_1}$) and covariance matrix accounting for the correlations between the components ($C_{H_1}$) are given by:
\begin{eqnarray}
\label{eq:mean_h1}
\bar{\xi}_{H_1}\left(i\right) & = & \frac{1}{N_s}\sum_{n=1}^{N_s} \xi_{H_1,n}\left(i\right), \\
\label{eq:cov_h1} C_{H_1}\left(j, k\right) & = & \frac{1}{N_s}\sum_{n=1}^{N_s} \big( \xi_{H_1,n}\left(j\right) - \bar{\xi}_{H_1}\left(j\right)\big) \times \\
& & ~~~~~~~~~~\big( \xi_{H_1,n}\left(k\right) - \bar{\xi}_{H_1}\left(k\right)\big). \nonumber
\end{eqnarray}
As before, $N_s$ is the total number of simulations (10,000) used to compute these estimators.

\subsection{The discriminator}
\label{subsec:discriminator}

The signal vectors defining the $H_0$ and $H_1$ hypotheses ($\xi_{H_0}$ and $\xi_{H_1}$, respectively) contain all the required information to distinguish between these two different scenarios. However, a practical way to add together all this information is required (each vector has $n_c$ components). There are different possibilities, such as building a $\chi^2$. However, we prefer to adopt a mechanism that provides an optimal\footnote{Fisher discriminant is proved to be optimal from the point of view
of adding the information through linear combinations~\citep[see, for instance,][]{cowan98}.} way to combine this information in the sense of obtaining the largest separation between the two hypotheses: the Fisher discriminant~\citep{fisher36}. The reader can find applications of the Fisher discriminant related to CMB Gaussianity studies in several works~\citep[e.g.,][]{barreiro01b,martinezgonzalez02}. The Fisher discriminant applied to $N$ signals corresponding to the $H_0$ hypothesis, provides a set of $N$ numbers ($\tau_{H_0}$) where all the information available for the null hypothesis $H_0$ (i.e., $\xi_{H_0}$, $\bar{\xi}_{H_0}$ and $C_{H_0}$) has been optimally combined. To construct this combination, the overall properties of the alternative hypothesis $H_1$ (i.e., $\bar{\xi}_{H_1}$ and $C_{H_1}$) are also taken into account. Analogously, the Fisher discriminant applied to $N$ signals following the $H_1$ hypothesis provides a set of $\tau_{H_1}$, that are built from the information related to the $H_1$ hypothesis 
(i.e., $\xi_{H_1}$, $\bar{\xi}_{H_1}$ and $C_{H_1}$), and the overall information related to the null hypothesis $H_0$ (i.e., $\bar{\xi}_{H_0}$ and $C_{H_0}$).

More specifically~\citep[see, for instance,][]{martinezgonzalez02}, for a given simulation $n$, the $\tau_{H_0}$ and $\tau_{H_1}$ quantities are given by:
\begin{eqnarray}
\label{eq:fisher_h0}
\tau_{H_0} & = & \big( \bar{\xi}_{H_0} - \bar{\xi}_{H_1} \big)^t W^{-1} \xi_{H_0}     , \\
\label{eq:fisher_h1} \tau_{H_1} & = &  \big( \bar{\xi}_{H_0} - \bar{\xi}_{H_1} \big)^t W^{-1} \xi_{H_1},
\end{eqnarray}
where $\left( . \right)^t$ denotes standard matrix transpose, and the matrix $W$ is obtained as $W = C_{H_0} + C_{H_1}$.

\section{Results}
\label{sec:results}

In this Section we present the results of applying the previously described methodology to CMB simulations. We have performed simulations that are compatible with the cosmological model determined by the analysis of the WMAP data~\citep{komatsu09}. The determination of the radial profiles in the temperature and (E-mode) polarization maps is performed at $\nside=64$, since only angular scales larger than 1 degree are considered. As mentioned in the previous Section, 10,000 simulations have been used to estimate the mean value of the signal vectors ($\xi_{H_0}$ and $\xi_{H_1}$), that contain the cross-correlation between the profiles $\mu_T\left(\x,\theta\right)$ and $\mu_E\left(\x,\theta\right)$, as well as the covariance matrices $C_{H_0}$ and $C_{H1}$ defining the correlation between the components of these vectors.

One thousand additional simulations have been used to calculate the distribution of the Fisher discriminants $\tau_{H_0}$ and $\tau_{H_1}$. We have studied the power of the proposed methodology to distinguish between the null ($H_0$) and the alternative ($H_1$) hypotheses, for different instrumental noise levels in the polarization maps ($\sigma_E$). In particular, we have studied in 
detail three scenarios corresponding to an ideal instrument ($\sigma_E \equiv 0$), to the QUIJOTE experiment~\citep[$\sigma_E \approx 0.3\mu$K per square degree, see][]{rubinomartin10}, and to the ESA Planck satellite~\citep[$\sigma_E \approx 1\mu$K per square degree, see][]{tauber10}.

In Figure~\ref{fig:fisher} we plot the distribution of the Fisher discriminants $\tau_{H_0}$ (solid blue lines) and $\tau_{H_1}$ (dot-dashed red lines) for these three cases: the ideal noise-free experiment is represented in the left panel, the output for the QUIJOTE experiment is provided in the middle panel, and, finally, the case for the Planck satellite is shown in the right panel. The vertical lines indicate the mean value for each distribution. At a power of the test $(1 - \beta) = 0.5$, the significance levels $\alpha$ are: 0.008 for the ideal experiment, 0.014 for the QUIJOTE experiment, and 0.069 for the Planck satellite.

\begin{figure}
\includegraphics[width=8cm,keepaspectratio]{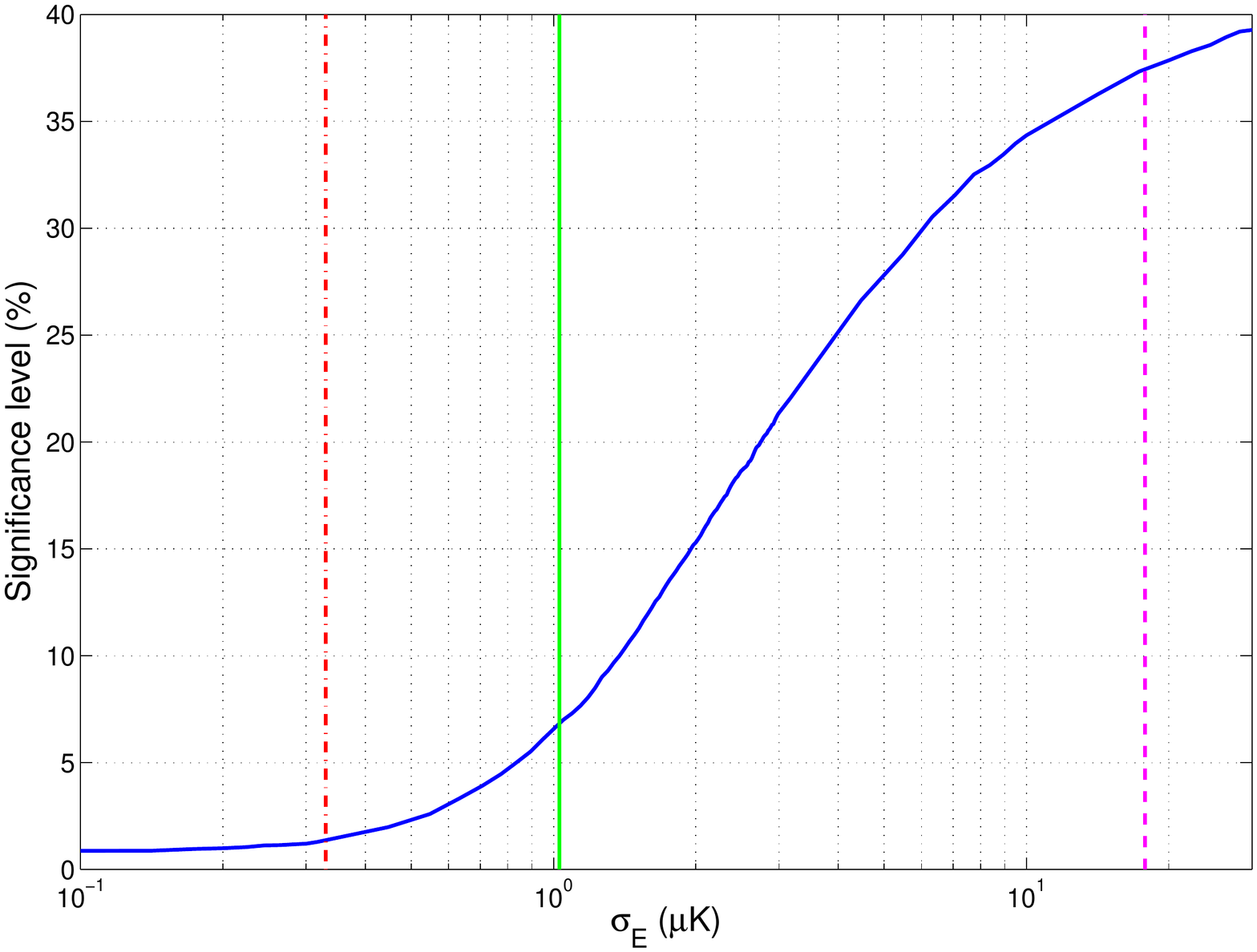}
\caption{\label{fig:s2n}Significance level to reject the $H_1$ hypothesis (at a power of the test of 0.5), as a function of the
instrumental noise level in the polarization map (given in $\mu$K per square degree). 
The vertical lines from left to right indicate the noise levels for QUIJOTE, Planck and WMAP5.}
\end{figure}

A more complete picture of the significance level to discriminate between the $H_0$ and $H_1$ hypotheses
is given in Figure~\ref{fig:s2n}, where the significance level (for a power of the test of 0.5) is shown as a function of the 
instrumental noise level $\sigma_E$. 
The vertical lines from left to right indicate the noise levels for QUIJOTE, Planck and WMAP5.

The previously estimated significance levels have been calculated for TE correlations \emph{given} that the temperature was anomalous. Therefore we can denote them as $P(TE | T)$. However, we can use both, TE and T, in order to discriminate between the null and alternative hypotheses. Hence we have $P(T,TE) = P(T) P(TE | T)$. We will set $P(T)=0.018$ since this is a very robust and conservative estimation for the p-value of the CS in temperature~\citep[see][for details]{cruz07a}. In this way, the $P(T,TE)$ significance levels (in percentage) are found to be 0.014\% for an ideal noise free experiment, 0.025\% for the QUIJOTE telescope and 0.12\% for Planck.

\section{Conclusions}
\label{sec:final}

The CMB polarization signature of the CS is proposed to distinguish between the possibility that it is just a rare fluctuation from
the Gaussian inflationary scenario (null hypothesis) or that it is due to the gravitational effect produced by a non-standard cosmological model, as for example the cosmic texture model (alternative hypothesis).
Obviously, cosmic textures are not the only physical process generating secondary anisotropies
via the non-linear integrated Sachs-Wolfe effect. For instance, a very large void in the
large scale structure, could generate |at least qualitatively speaking| a similar effect.
However, as many works have already indicated~\citep[e.g.,][]{cruz08,smith10,granett09,bremer10},
the void hypothesis is very unlikely. On the contrary,
cosmic textures have proven to be a plausible explanation~\citep{cruz07a,cruz08}.

Whereas polarization alone is not enough to discriminate between the two hypotheses, the TE cross-correlation provides
a significant signal. In the case that the null hypothesis is
correct, one would expect a significant cross-correlation signal with an amplitude
corresponding to that of the largest spot in the
temperature sky. On the contrary, if the alternative hypothesis is
true, no additional polarization (and thus no cross-correlation) signal would
be expected from the gravitational effect of the texture collapse. In
this latter case, the only expected TE signal would be the one corresponding
to a random inflationary fluctuation.
      
The test proposed in this paper makes use of the Fisher discriminant constructed from all possible TE cross-correlation combinations formed from the temperature and polarization profiles at the position of a CS-like feature. In the best case of an ideal noise-free polarization experiment the null hypothesis for the CS TE signal can be rejected at a significance level of 0.8\%. 
For the case of QUIJOTE and Planck this result becomes 1.4\% and 6.9\%, respectively.

Finally, we may wonder about the probability at which the null hypothesis can be rejected by taking into account both the temperature
and polarization information of the CS. Considering that in the inflationary scenario the probability of having a temperature as extreme as the one measured for the CS is 1.8\%~\citep{cruz07a}, then the combination of this probability with the one found in this work using polarization information, would provide a significance of 0.014\%, 0.025\% and 0.12\% for the ideal, QUIJOTE and Planck experiments, respectively.

\section*{Acknowledgements}
We acknowledge partial financial support
from the Spanish Ministerio de Ciencia e Innovaci{\'o}n project
AYA2007-68058-C03-02. PV also thanks financial support from
the \emph{Ram\'on y Cajal} programme. 
MT thanks the Physikalisch-Meteorologisches Observatorium 
Davos/World Radiation Center (PMOD/WRC) for having provided 
him facilities to carry out the investigation.
The HEALPix package was 
used throughout the data analysis \citep{gorski05}.

\label{lastpage}

\end{document}